\documentclass [12pt,a4paper]{article}                  
\usepackage{times}

\DeclareFontFamily{OT1}{times}{}
\DeclareFontShape {OT1}{times}{m }{n }{ <-> ptmr }{}
\DeclareFontShape {OT1}{times}{bx}{n }{ <-> ptmb }{}
\DeclareFontShape {OT1}{times}{m }{it}{ <-> ptmri}{}
\DeclareFontShape {OT1}{times}{bx}{it}{ <-> ptmbi}{}
\usepackage{amsmath}
\usepackage{amsfonts}
\usepackage{amssymb}
\usepackage{latexsym}

\numberwithin{equation}{section}

\newcommand{\HU}{\text{H}} 
\newcommand{\sgn}{\operatorname{sgn}} 
\newcommand{\BRA}{\langle\kern -.2em\langle} 
\newcommand{\KET}{\rangle\kern -.2em\rangle} 
\newcommand{\DEF}{:=}                 
\begin{document}

\title{\bf\vspace{-2.5cm}      Distributions in spherical coordinates
                           with applications to classical electrodynamics}

\author{
         {\bf Andre Gsponer}\\
         {\it Independent Scientific Research Institute}\\ 
         {\it Oxford, OX4 4YS, England}
       }

 \date{Eur. J. Phys. {\bf 28} (2007) 267--275}

\maketitle

\begin{abstract}

A general and rigorous method to deal with singularities at the origin of a polar coordinate system is presented.  Its power derives from a clear distinction between the radial distance and the radial coordinate variable, which makes that all delta-functions and their derivatives are automatically generated, and insures that the Gauss theorem is correct for any distribution with a finite number of isolated point-like singularities.

The method is applied to the Coulomb field, and to show the intrinsic differences between the dipole and dimonopole fields in classical electrodynamics.

In all cases the method directly leads to the general expressions required by the internal consistency of classical electrodynamics.

\end{abstract}	

\section{Introduction} 
\label{int:0}

How to deal in a consistent and systematic manner with singularities arising at the origin of a polar coordinates system is a recurring question in the teaching of classical electrodynamics \cite{FRAHM1983-,BLIND2003-, HU---2004-, LEUNG2006-} as well as in research \cite{TANGH1962-, GSPON2004C,GSPON2006C}.

The standard solution is pragmatic: it consists of introducing an appropriate delta-function whenever the derivative of an expression that is singular at the origin is a distribution rather than zero, i.e., when the derivative is zero everywhere except at the origin.  Typical examples in three dimensional space are the formulas for the divergence of the Coulomb field
\begin{equation} \label{int:1}
                          \vec \nabla \cdot \frac{\vec x}{|\vec x|^3} 
                   = \frac{1}{|\vec x|^2}\delta(|\vec x|)
                   =   4\pi \delta^3(\vec x),
\end{equation}
and the Laplacian of the Coulomb potential
\begin{equation} \label{int:2}
                             \Delta \frac{1}{|\vec x|}
                     =    \vec \nabla \cdot \vec \nabla \frac{1}{|\vec x|} 
                     =  - 4\pi \delta^3(\vec x) .
\end{equation}
While these formulas are perfectly correct, as can be verified in a number of ways, it would be desirable to have a method such that the ``delta-functions'' appear at the right place without having to remember when this or that expression yields a distribution instead of zero.  Moreover, one would like to have a technique that is general and mathematically rigorous, and whose application is straightforward.

    Indeed, the only thing distribution theory says about an expression that is zero everywhere except at a single point derives from the following theorem: \emph{A distribution which has its support only in one point, say the origin, is a linear combination of the delta-function and its derivatives up to a certain order}.\footnote{For a proof of that theorem, see \cite[p.~784]{COURA1962-} or \cite[p.~443]{CHOQU1982-}.}  Therefore, while this theorem gives a rigorous justification to the standard method consisting of introducing the delta-functions ``by hand,'' and of indirectly finding their coefficients by a way or another, it does not provide a rule specifying how to formulate an initial expression as a distribution so that there is no ambiguity in differentiating it in order to get the correct result in a straightforward manner.

   In this paper we present such a method.  It is directly applicable to the elementary case of an isolated point-like singularity of a scalar or vector field over $\mathbb{R}^3$, which therefore may be considered as being located at the origin of a spherical coordinate system.  The method can immediately be extended to $n$-dimensional spaces and fields containing a finite number of isolated point singularities.  But it cannot easily be extend to topologically more complicated singularities such as, for example, the Dirac or Schwinger magnetic monopoles which have a potential with a line-like singularity.  For similar reasons the distributions considered in this paper will always be continuous in the angular variables.

   The essence of the method consists of writing the field under consideration in such a way that there is no ambiguity with regards to how to calculate its derivatives at every point, including the origin where it may be singular. This is made possible by a clear distinction between the radial distance and the radial coordinate variable, a distinction that was first made by Tangherlini in the context of special and general relativity \cite[p.~511--513]{TANGH1962-}, which led him to use the sign-function $\sgn(x)$ to specify how to differentiate at the origin.  This idea was rediscovered by the author, and others who used it mainly as an ansatz \cite{BLIND2003-, HU---2004-}, whereas it leads in fact to a rigorous  method.  It is explained and justified in Sec.~\ref{dis:0}, where an effort is made to be as correct as possible, but without pretending to a rigor such that professional mathematicians would be fully satisfied.  In particular, as little mathematical background as possible is used, and the interested reader is referred to the numerous publications in which the theory of distributions is presented.\footnote{For a comprehensive introduction to distribution theory see, e.g., \cite[p.~766--798]{COURA1962-} or \cite[p.~423--541]{CHOQU1982-}, and for a concise modern presentation \cite{SCHUC1991-}.}

In Sec.~\ref{gau:0} it will be shown that Gauss's theorem formulated in polar coordinates is correct even if there is a singularity at the origin, provided the method developed in this paper is used to calculate the derivatives.  This is an important consistency check since Gauss's theorem enables to move between the differential and the integral formulations of electrodynamics.

In Sec.~\ref{mon:0} the method is applied to the most simple non-trivial case: the monopole singularity of a Coulomb or Newton point charge.  It will be found that an additional singular field appears along side the regular field when it is derived from the potential.  This additional contribution is essential to insure the local conservation of the electromagnetic charge-current density.

In Secs.~\ref{dip:0} and \ref{dim:0} the potentials, fields, and sources of the dipole and dimonopole singularities are studied.  It will be shown that despite that the dimonopole is a scalar-potential-singularity, and the dipole a vector-potential-singularity, their fields are identical, except at the origin where the two singularities have different delta-function-like fields.  It will also be seen that while the derivation of this important result is generally somewhat indirect in standard textbooks, it derives rigorously from a straightforward application of the present method.

\section{Distributions in spherical coordinates} 
\label{dis:0}

In electrodynamics and other areas of physics one is often led to calculating integrals of the form
\begin{equation}\label{dis:1}
  \BRA D | T \KET \DEF \iiint\limits_{\mathbb{R}^3} d^3\Omega~   D(\vec{x}) T(\vec{x}),
\end{equation}
where $T(\vec{x})$ is a well behaved function (i.e., indefinitely differentiable and vanishing outside a bounded region) and $D(\vec{x})$ a function which may be singular, that is infinite, at some point.  For example $T(\vec{x})$ could be the velocity of a stream of charges, $D(\vec{x})$ an electromagnetic field, and the integral the electromagnetic force on that stream.  In the language of distribution theory, $T$ is called a test function, $D$ a distribution, and the inner product $\BRA D | T \KET$ the value of $D$ on $T$.

If the distribution is singular at a point $\vec{x}_0$ it is natural to take this point as the origin of a polar coordinate system $\vec{x}-\vec{x}_0=\vec{r}(r,\theta,\phi)$.  However, unless one is very careful, the use of such a coordinate system can lead to ambiguities.  For instance, the standard practice is to rewrite Eq.~\eqref{dis:1} as follows
\begin{equation}\label{dis:2}
  \BRA D | T \KET = \int_{0}^{2\pi}d\phi \int_{0}^{\pi}d\theta \sin \theta
                    \int_{0}^{\infty}dr~ r^2 D(r,\theta,\phi) T(r,\theta,\phi),
\end{equation}
where the radial coordinate
\begin{equation}\label{dis:3}
     r  = |\vec{r}| \DEF \sqrt{\vec{r}\cdot\vec{r}},
\end{equation}
is a positive number. This leads to two difficulties. First, as $r\in\mathbb{R}^+$ and the singularity is at $r=0$, that is at the origin of a half-space, a number of basic results of distribution theory require some adaptation since they are generally formulated for distributions defined over some variable $x\in\mathbb{R}$ and for singularities at $x=0$.  Second, again as $r\geqslant 0$, one has to carefully distinguish between the coordinate ``$r$,'' whose differential $dr$ is part of the integration element, and its magnitude ``$|r|$,'' which may appear in the definition of some particular distribution --- which for clarity should therefore be written $D(r,\theta,\phi, |r|)$.  

   To better see why this distinction is necessary, let us rewrite Eq.~\eqref{dis:2} using another spherical coordinate system, parametrized in such a way that the radial coordinate is now a signed real number $x$, i.e.,
\begin{equation}\label{dis:4}
  \BRA D | T \KET = \int_{0}^{\pi}d\phi \int_{0}^{\pi}d\theta \sin \theta 
     \int_{-\infty}^{+\infty}dx~ x^2  D(x,\theta,\phi,|x|) T(x,\theta,\phi).
\end{equation}
In this case the whole set of standard distribution theory theorems can directly  be used, and there is less temptation to confuse the coordinate $x$ with its absolute value $|x|$.  Indeed, while $|x|$ is continuous at $0$, its derivative is discontinuous at this point since it goes from $+1$ to $-1$, or vice versa, when $x$ goes through $0$. For this reason the second derivative of $|x|$ gives rise to a $\delta$-function at the origin.  The absolute value $|x|$ is therefore \emph{not} differentiable at the origin, which is why it must be interpreted as a distribution, whereas the coordinate $x$ is a regular variable for which there is no problem at the origin.  In fact, the same is true for $|r|$ and $r$ in the standard parametrization \eqref{dis:2}, except that in this case these differences are somewhat hidden since the integration over the negative part of the $x$ axis is replaced by an extension of the $\phi$ integration range from $[0,\pi]$ to $[0,2\pi]$.

   Consequently, to work with distributions in polar coordinates it is necessary to insure that all references to a distance or magnitude $|x|$ are made explicit.  With the spherical parametrization \eqref{dis:4}, this is best done by introducing the sign function $\sgn(x)$ whose definition and relation to Dirac's $\delta$-function are 
\begin{equation} \label{dis:5}
   \sgn(x) = 
         \begin{cases}
         -1     &   x < 0,\\
         0      &   x = 0,\\
         +1     &   x > 0,
         \end{cases}
   ~~~  ~~~ \text{and} ~~~ ~~~
     \frac{d}{dx}\sgn(x) = 2 \delta(x).
\end{equation}
For any occurrence of $|x|$ we can then write
\begin{equation}\label{dis:6}
                   |x| = x\sgn(x),  ~~~ ~~~ 
       \frac{d}{dx}|x| = \sgn(x),  ~~~ ~~~ 
   \frac{d^2}{dx^2}|x| = 2 \delta(x), ~~~ ~~~ \text{etc}., 
\end{equation}
where the rule $x\delta(x)=0$ was used to go from $|x|$ to $d|x|/dx$.

   However, the polar parametrization \eqref{dis:2} is far more frequently used than the non-standard parametrization \eqref{dis:4}.  In this case we may still use the sign function $\sgn(x)$ and write $|r|=\sgn(r)$ even though $r$ never takes a negative value.  But this leads to practical difficulties since the radial integration will no more be from $-\infty$ to $+\infty$, but from $0$ to $+\infty$, so that, for example, the normalization of Dirac's $\delta$-function would have to be $1/2$ instead of $1$ to compensate for the factor $2$ in Eq.~\eqref{dis:5}.  Since this may lead to confusion, and basically consists of formally ``splitting'' the $\delta$-function at $r=0$, one may just as well ``split'' the sign function and define a generalized function $\Upsilon(r)$ such that
\begin{equation} \label{dis:7}
   \Upsilon(r) \DEF 
         \begin{cases}
         \text{undefined}   &   r < 0,\\
                      0     &   r = 0,\\
                     +1     &   r > 0,
         \end{cases}
   ~~~  ~~~ \text{and} ~~~ ~~~
     \frac{d}{dr}\Upsilon(r) = \delta(r).
\end{equation}
This function should not be confused with Heaviside's step function $\HU(r)$, which is 0 for $r<0$ and undefined at $r=0$, whereas $\Upsilon(r)$ is undefined for $r<0$, and equal to $0$ at $r=0$ to be equivalent to $\sgn(x)$ at $x=0$.

    Therefore, for any occurrence of $|r|$ in the standard parametrization \eqref{dis:2} we shall write 
\begin{equation}\label{dis:8}
                   |r| = r\Upsilon(r),  ~~~ ~~~ 
       \frac{d}{dr}|r| =  \Upsilon(r),  ~~~ ~~~ 
   \frac{d^2}{dr^2}|r| =  \delta(r), ~~~ ~~~ \text{etc}. 
\end{equation}
In particular, for any occurrence of the radius vector $\vec{r}$ we shall write
\begin{equation}\label{dis:9}
     \vec{r} = r \Upsilon(r) \vec{u}(\theta,\phi), 
\end{equation}
where $\vec{u}$ is the unit vector in the direction of $\vec{r}$. 

Finally, we shall make use of the property  
\begin{equation}\label{dis:10}
  \int_0^\infty dr~ \Upsilon(r)T(r) =  \int_0^\infty dr~ T(r),
\end{equation}
whose consistency with Eq.~\eqref{dis:7} can be verified by integrating by part the left-hand side, i.e.,
\begin{equation}\label{dis:11}
   \Upsilon(r)T^{(-1)}(r)\Bigr|_0^\infty - \int_0^\infty dr~ \delta(r)T^{(-1)}(r)
                                   = \int_0^\infty dr~ T(r),
\end{equation}
where we made use of the specification $\Upsilon(0)=0$, as well as of the equation
\begin{equation}\label{dis:12}
  \int_0^\infty dr~ \delta(r)T(r) = T(0).
\end{equation}

    With these rules everything related to the use of a polar coordinate system is rigorously taken care of.  What remains to make sure is that the generalized functions $D(r,\theta,\phi,|r|)$ are really distributions, i.e., that their inner products $\BRA D | T \KET$ are converging for any test function $T(r) \in \mathcal{C}^\infty$.  This is the case of the distributions considered in this paper, which have singularities of type $r^{-n}$ with $n \geq 1$ at $r=0$.  Indeed, we must carefully distinguish between the classical functions $r^{-n}$ and their derivatives $-nr^{-n-1}$, which are defined only for $r >0 $, and the corresponding distributions which are defined for all $r \geq 0$.  This is done by defining these functions as limits of sequencies of distributions \cite[p.~51]{SCHUC1991-}, i.e., 
\begin{equation}\label{dis:13}
      \frac{1}{r^n} \DEF \lim_{\epsilon \rightarrow 0} \frac{1}{r^n} \HU(r-\epsilon).
\end{equation}

\section{Gauss's theorem in polar coordinates}
\label{gau:0}

In order to illustrate the power and generality of the method just explained, we give in this section an elementary proof that Gauss's theorem is true in the distributional sense for any weakly convergent scalar- or vector-valued distribution $F(\vec{r})$ which may have a singularity at $r=0$.  That is, for any finite simply-connected 3-volume $\Omega$ bounded by the 2-surface $\Sigma = \partial \Omega$, we prove that
\begin{equation}\label{gau:1}
 \iint\limits_{\partial\Omega} d^2\Sigma~ F(\vec{r}) = 
\iiint\limits_{        \Omega} d^3\Omega~ \vec\nabla F(\vec{r}).
\end{equation}
Referring to Eq.~\eqref{dis:1} we have here $T=1$, and $D=\vec\nabla F(\vec{r})$ is weakly convergent because it is a combinations of partial derivatives of the distribution $F(\vec{r})$. 

   Since $F(\vec{r})$ is supposed to be finite and differentiable everywhere except at $r=0$, the theorem is true by any standard proof for any volume which does not contain the origin.  This enables to proceed as in any standard analysis of point-like singularities, namely to surround the origin by a small sphere and prove the theorem for the leading singularity.  However, for simplicity, we postpone to a more complete proof the cases where the origin is at the surface of the volume $\Omega$, and the one where $\Omega$ is multiply connected.

   From now one the volume $\Omega$ is therefore an ordinary 3-ball of radius $R$, $\partial\Omega$ is its surface, and the function  $F(\vec{r})$ is the leading singularity.  If $F(\vec{r})$ is simply a scalar function $f(r)$, which does not depend on the angular variables, the theorem reduces after angular integration to the identity $0=0$.  This remains so after multiplying $f(r)$ by a constant vector, so that to deal with non trivial singularities we must consider expressions containing the vector $\vec{r}$.  The most simple such a singularity, expressed according to Eq.~\eqref{dis:9}, is
\begin{equation}\label{gau:2}
  F(\vec{r}) = \frac{\vec{r}}{r^n} = \frac{\vec{u}}{r^{n-1}}\Upsilon(r),
\end{equation}
which is unbounded at $r=0$ for $n > 1$.  We can now operate with $\vec\nabla$ on $\Upsilon(r)$, and make use of Eq.~\eqref{dis:10} to remove $\Upsilon(r)$ where it has become un-necessary, so that Eq.~\eqref{gau:1} becomes
\begin{equation}\label{gau:3}
 \iint\limits_{\partial\Omega} d^2\Sigma~ F(\vec{r}) = 
\iiint\limits_{        \Omega} d^3\Omega~ \vec\nabla \frac{\vec{u}}{r^{n-1}} +
\iiint\limits_{        \Omega} d^3\Omega~ \vec{u} \frac{\vec{u}}{r^{n-1}}
                                                  \delta(r).
\end{equation}

 For the 3-ball, we have
\begin{equation}\label{gau:4}
  d^2\Sigma = d\phi\, d\theta \sin\theta\, r^2 \vec{u}
               ~~~~ ~~~~ \text{and} ~~~~ ~~~~ 
  d^3\Omega = d\phi\, d\theta \sin\theta\, r^2 dr .
\end{equation}
Thus, after an elementary calculation, the three terms appearing in $\eqref{gau:3}$ become
\begin{gather}
    \iint\limits_{\partial\Omega} d^2\Sigma~ F(\vec{r}) =
    4\pi  r^2 \vec{u}\frac{\vec{u}}{r^{n-1}} \Bigr|^{R} = 
    4\pi R^{3-n},
\label{gau:5} \\
    \iiint\limits_{    \Omega} d^3\Omega~ \vec\nabla \frac{\vec{u}}{r^{n-1}} =
    4\pi \int_{0}^{R} dr~ (3-n)r^{(2-n)} =
    4\pi   r^{3-n} \Bigr|_{r \rightarrow 0}^{R} ,    
\label{gau:6}\\
\iiint\limits_{      \Omega} d^3\Omega~ \frac{1}{r^{n-1}}\delta(r)=
    4\pi \int_{0}^{R} dr~ r^2 \frac{1}{r^{n-1}} \delta(r)=
    4\pi   r^{3-n} \Bigr|^{r \rightarrow 0} ,
\label{gau:7}
\end{gather}
where the lower limits of the volume integrals have not been taken in order to show the impact of having used Eq.~\eqref{dis:9}.  Indeed, by comparing $\eqref{gau:6}$ and $\eqref{gau:7}$, we see that the second term on the right hand side of $\eqref{gau:3}$ has the remarkable effect of removing the divergent term at $r \rightarrow 0$, so that the result is equal to the surface integral $\eqref{gau:5}$, and, moreover, in the case $n=3$, of correcting equation $\eqref{gau:6}$ so that instead of zero it gives the same value as $\eqref{gau:5}$.  In other words, that second term ``repairs'' the usual formulation of Gauss's theorem in such a way that it becomes true for all singularities of the type~$\eqref{gau:2}$.  From there on it is easy to generalize the proof to more complicated functions and formulations, so that Gauss's theorem in its various forms is true in polar coordinates, even if there are singularities at the origin, provided the rules given in Eqs.~\eqref{dis:7} to \eqref{dis:12} are followed.

\section{Monopole singularity} 
\label{mon:0}

When considering the $1/r$ potential of a Newton or Coulomb field, the straightforward application of Eq.~\eqref{dis:8} is ambiguous because of the algebraic identity
\begin{equation} \label{mon:1}
       \frac{1}{|r|} = \frac{|r|}{r^2}.
\end{equation}
However, in order to obtain the correct form of the field, for which the application of Eq.~\eqref{dis:9} is non-ambiguous, it is evident that the potential of a point-charge is  
\begin{equation} \label{mon:2}
       \varphi_m(\vec r) \DEF e\frac{1}{r}\Upsilon(r),
\end{equation}
so that the Coulomb field is
\begin{equation} \label{mon:3}
       \vec E_m(\vec r) = -\vec \nabla \varphi_m
                        =  e\frac{\vec r}{r^3}\Upsilon(r) 
                        -  e\frac{\vec r}{r^2}\delta(r). 
\end{equation}
The rationalized source-charge distribution is then
\begin{equation} \label{mon:4}
       4\pi \rho_m(\vec r) =    \vec \nabla \cdot \vec E_m 
                           =  e \frac{1}{r^2} \delta(r),
\end{equation}
which upon integration yields the {charge} of the source
\begin{equation}\label{mon:5}
     q = \BRA \rho_m | 1 \KET = \iiint d^3\Omega ~ \rho_m(\vec r) = e.
\end  {equation}

There is however an important difference between Eq.~\eqref{mon:3} and the usual textbook expressions for the Coulomb field: the additional $\delta$-like term on the right of $\eqref{mon:3}$.  This requires a careful analysis because the physical interpretation of this term is different depending on the physical significance given to the evaluation of $\vec E_m$ on a test function, i.e., to the inner product $\BRA \vec E_m | T \KET$, and to the expression $\vec E_m(\vec r)$ itself.  For instance:

\begin{enumerate}

  \item When calculating $\BRA \vec E_m | T \KET$ it is immediately seen that the $\delta$-like term does not contribute to the radial integral because the $r^2$ factor in $d^3\Omega$ leads to a  $r\delta(r)$ product which is zero.  Therefore, from the perspective of distribution theory, the usual Coulomb field $\vec{E}_C=e\vec{r}/r^3$ and the field $\vec{E}_m(\vec r)$ are equivalent, that is corresponding to the same distribution.  However, the inner product $\BRA \vec{E}_m | T \KET$ has no physical meaning because it has no invariant significance, contrary to the charge, Eq.~\eqref{mon:4}.  The difference between $\vec{E}_C$ and $\vec{E}_m$ has therefore to be evaluated from a physical point of view. 

\item The $\delta$-like term in Eq.~\eqref{mon:3} is a solution of the homogeneous equation $\vec\nabla \cdot\vec E_m(\vec r)=0$, which means that it can always be added to a solution of the inhomogeneous equation \eqref{mon:4}.  But this does not imply that this singular term can always be discarded.  For example, in the relativistic case where the potential and fields of an arbitrarily moving charge are considered, i.e., in the  Li\'enard-Wiechert case, the additional $\delta$-like field is absolutely necessary to insure local charge conservation \cite{GSPON2006C}.

\end{enumerate}

In summary, the derivation of the Coulomb charge distributions \eqref{mon:4} from the potential \eqref{mon:2} has lead to the usual charge distribution, i.e., Eqs.~\eqref{int:1} or \eqref{int:2}, whereas the Coulomb field distribution turned out to have the more general form \eqref{mon:3} required by the internal consistency of classical electrodynamics.

\section{Dipole singularity} 
\label{dip:0}

The dipole singularity, which through extensive experimental verification is found to very precisely characterize the intrinsic magnetic dipole moment of elementary particles such as the electron, is given by the {vector} potential
\begin{equation}\label{dip:1}
   \vec A_d(\vec r) \DEF
             \frac{\vec{\mu} \times \vec{r} }{r^3}\Upsilon(r),
\end{equation}
where $|\vec{\mu}\,|$ has the dimension of a charge times a length. The calculation of the magnetic field strength gives
\begin{equation}\label{dip:2}
    \vec H_{d}(\vec r) =  \vec\nabla \times \vec A_d =
        \Bigl( 3\frac{\vec  r}{r^5}(\vec\mu\cdot\vec r)
              - \frac{\vec\mu}{r^3} \Bigr)  \Upsilon(r)
 +        \frac{\vec{r} \times (\vec\mu\times\vec{r})}{r^4}\delta(r).
\end{equation}
The first term in this expression is well-known, but the one with a $\delta$-function is rarely mentioned in textbooks. However, when integrated over 3-space, this second term gives the contribution \cite[p.~184]{JACKS1975-}
\begin{equation}\label{dip:3}
      + \iiint d^3\Omega ~
       \delta(r)~ \frac{\vec{r} \times (\vec\mu\times\vec{r})}{r^4}
     = \frac{8\pi}{3} \vec\mu ,
\end{equation}
which is essential in calculating the hyperfine splitting of atomic states \cite{JACKS1977-}. Thus, as in Eq.~\eqref{mon:3} for the monopole singularity,  we have in Eq.~\eqref{dip:2} a physically essential $\delta$-like contribution, which was obtained by a straightforward application of the method explained in Sec.~\ref{dis:0}.

   We can now calculate the sources.  As expected, the magnetic charge density is zero
\begin{equation}\label{dip:4}
      4\pi \rho_{d}(\vec r) = \vec\nabla \cdot \vec H_{d}(\vec r) = 0  ,
\end{equation}
while the rationalized current density is
\begin{equation}\label{dip:5}
    4\pi \vec j_{d}(\vec r) = \vec\nabla \times \vec H_{d}(\vec r)
        =   3 \frac{\vec{\mu} \times \vec{r} }{r^4} \delta(r) . 
\end{equation}
Using this current density we can now calculate the {magnetic moment} by means of the standard expression \cite[p.~181]{JACKS1975-} to get
\begin{equation}\label{dip:6}
            \vec m = 
 \frac{1}{2} \iiint d^3\Omega ~ \vec r \times \vec j_{d}(\vec r)
                   =  \vec \mu .
\end{equation}
Therefore, although there are actually no ``circulating currents'' in the point-like distribution $\eqref{dip:5}$, the magnetic moment calculated with the formula derived for a localized current distribution gives the correct answer.

\section{Dimonopole singularity} 
\label{dim:0}

The dimonopole singularity corresponds to the field produced by two magnetic (or electric) monopoles of opposite charge $\pm q$ separated by an infinitesimal distance $|\vec \lambda|$. The potential for such a field is therefore the scalar expression
\begin{equation}\label{dim:1}
   \varphi_{dm}(\vec r) \DEF \frac{q}{|\vec{r}\,|}
                           - \frac{q}{|\vec r + \vec\lambda|}.
\end{equation}
At large distance, or at vanishingly small separation $\vec \lambda$, we can take for this potential the first term of the Taylor development, i.e.,
\begin{equation}\label{dim:2}
            \varphi_{dm}(\vec r)
            \approx q \frac{1}{r^3} (\vec\lambda \cdot \vec r) \Upsilon(r).
\end{equation}
The field strength is then
\begin{equation}\label{dim:3}
    \vec H_{dm}(\vec r) = - \vec{\nabla} \varphi_{dm} =
    \Bigl( 3\frac{\vec r}{r^5}(\vec\mu\cdot\vec r)
          - \frac{\vec\mu}{r^3} \Bigr)  \Upsilon(r)
  -  \frac{\vec{r}(\vec\mu\cdot\vec{r})}{r^4} \delta(r),
\end{equation}
where we have defined
\begin{equation}\label{dim:4}
                 \vec\mu = q \vec\lambda  .
\end{equation}
Expression $\eqref{dim:3}$ is remarkably similar to the corresponding expression $\eqref{dip:2}$ for an intrinsic dipole, and it can be seen that the difference between a dipole and a dimonopole field is entirely contained in the point-like singularity at the origin, i.e., 
\begin{equation}\label{dim:5}
    \vec H_{d}(\vec r) - \vec H_{dm}(\vec r)=  
         \frac{\vec{r} \times (\vec\mu\times\vec{r})}{r^4} \delta(r) +
         \frac{\vec{r}        (\vec\mu\cdot \vec{r})}{r^4} \delta(r) =
         \frac{\vec{\mu}}{r^2} \delta(r).
\end{equation}
As a result, when integrated over 3-space, the dimonopolar $\delta$-singular term in Eq.~$\eqref{dim:3}$ gives the contribution \cite[p.~141]{JACKS1975-}
\begin{equation}\label{dim:6}
       - \iiint d^3\Omega ~
         \frac{\vec{r}(\vec\mu\cdot\vec{r})}{r^4} \delta(r)
     = - \frac{4\pi}{3} \vec\mu, 
\end{equation}
which differs in sign and in magnitude from the corresponding expression $\eqref{dip:3}$ for an intrinsic dipole.  It is this difference which enables to conclude that the dipolar fields from distant stars are produced by magnetic dipoles, rather than by magnetic dimonopoles \cite{JACKS1977-}.

We can now calculate the sources.  As expected, the current density is zero
\begin{equation}\label{dim:7}
    4\pi \vec j_{dm}(\vec r) = \vec\nabla \times \vec H_{dm}(\vec r) = 0  ,
\end{equation}
while the rationalized charge density is
\begin{equation}\label{dim:8}
    4\pi \rho_{dm}(\vec r) = \vec\nabla \cdot \vec H_{dm}(\vec r)
        =   3 \frac{\vec{r} \cdot \vec{\mu} }{r^4} \delta(r) ,
\end{equation}
i.e., a distribution that is odd in $\vec{r}$ so that the total charge is zero, as it should be for a dimonopole. We can finally calculate the {first moment} of this charge density by means of the standard expression for a charge distribution \cite[p.~137]{JACKS1975-}. This gives
\begin{equation}\label{dim:9}
            \vec d = 
     \iiint d^3\Omega ~ \vec r \rho_{dm}(\vec r)
                 = \vec\mu  =  q\vec \lambda ,
\end{equation}
a result formally similar to Eq.~\eqref{dip:6} which was obtained by using the formula for the moment of a current distribution.  This illustrates again that despite the great similarity of their fields at a distance from the origin, the dipole and dimonopole singularities are in fact very different.

\end{document}